\documentclass[doublecol]{epl2}

\title{Normal heat conductivity in chains capable of dissociation}

\author{O. V. Gendelman\inst{1} \and A. V. Savin\inst{2}}
\institute{
  \inst{1} Faculty of Mechanical Engineering, Technion --
           Israel Institute of Technology,
           Haifa 32000, Israel\\
  \inst{2} Semenov Institute of Chemical Physics, Russian Academy
           of Sciences -- Moscow 119991, Russia
}

\abstract{
The paper suggests a resolution for recent controversy over convergence of heat conductivity
in one-dimensional chains with asymmetric nearest-neighbor potential. We conjecture that the
convergence is promoted not by the mere asymmetry of the potential, but due to ability
of the chain to dissociate. In other terms, the attractive part of the potential
function should approach a finite value as the distance between the neighbors grows.
To clarify this point, we study the simplest model of this sort -- a chain of linearly elastic
disks with finite diameter. If the distance between the disk centers exceeds their diameter,
the disks cease to interact. Formation of gaps between
the disks is the only possible mechanism for scattering of the oscillatory waves. Heat conduction
in this system turns out to be convergent. Moreover, an asymptotic behavior of the heat conduction coefficient
for the case of large densities and relatively low temperatures obeys simple Arrhenius-type law.
In the limit of low densities, the heat conduction coefficient converges due to triple disk
collisions. Numeric observations in both limits are grounded by analytic arguments.
In a chain with Lennard-Jones
nearest-neighbor potential the heat conductivity also saturates in a thermodynamic limit and
the coefficient also scales according to the Arrhenius law for low temperatures. This finding
points on a universal role played by the possibility of dissociation, as convergence
of the heat conduction coefficient is considered.
}
\begin{document}

\maketitle

Heat conduction in low-dimensional systems has attracted a lot of attention and has been a
subject of intensive studies \cite{LLP03,LiREV2012}. The main objective here is to derive from first
principles (on the atomic-molecular level) the Fourier law -- proportionality of the heat flux
to the temperature gradient $J=-\kappa \nabla T$, where $\kappa$ is the heat conduction coefficient.
To date, there exists quite extensive body of works devoted to the numerical modeling of the heat
transfer in the one-dimensional chains. Anomalous characteristics of this process are well-known
since celebrated work of Fermi, Pasta and Ulam  \cite{FPU}. In integrable systems (harmonic chain,
Toda lattice, the chain of rigid disks) the heat flux $J$ does not depend at all on the chain
length $L$, therefore, the thermal conductivity formally diverges. The underlying reason for that
is that the energy is transferred by non-interacting quasiparticles and therefore one cannot
expect any diffusion effects. Non-integrability of the system is a necessary but not  sufficient
condition to obtain the  convergent heat conduction coefficient. Well-known examples are
Fermi-Pasta-Ulam (FPU) chain \cite{LRP,LLP1,LLP2}, disordered harmonic chain \cite{RG,CL,D},
diatomic 1D gas of colliding particles \cite{D1,STZ02,GNY} and the diatomic Toda lattice \cite{H}.
In these non-integrable systems also have divergent heat conduction coefficient; the latter
diverges as a power function of length: $\kappa\sim L^\alpha$, for $L\rightarrow\infty$.
The exponent lies in  the interval $0<\alpha <1$.

On the other side, the 1D lattice with on-site potential can have finite conductivity.
The simulations had demonstrated the convergence of the heat conduction coefficient for
Frenkel-Kontorova chain \cite{HLZ98,SG03}, the chain with hyperbolic sine on-site potential \cite{TBSZ},
the chain with $\phi^4$ on-site potential \cite{HLZ00,AK00} and for the chain of hard disks of
non-zero size with substrate potential \cite{GS04}.
The essential feature of all these models is existence of an external potential modeling the
interaction with the surrounding system. These systems are not translationally invariant, and,
consequently, the total momentum is not conserved. In paper \cite{HLZ98} it has been suggested that
the presence of an external potential plays a key role to ensure the convergence of the heat
conductivity.  This hypothesis has been disproved in works \cite{giardina,savin2},
where it was shown that the isolated chain of rotators (a chain with a periodic potential
interstitial interaction) has a finite thermal conductivity.
Some recent studies \cite{ZZWZ12,CZWZ12} claim that the heat conduction is convergent
in some model chains with an asymmetric potential of interaction. In particular, this claim addressed
$\alpha$-FPU model and a chain of particles with Lennard-Jones (LJ) interaction. Results \cite{Den1,Den2}
deny these claims for $\alpha$-FPU model; however, they do not address the LJ chain. More detailed
results in recent paper \cite{SK13} indicate that the asymmetry alone is insufficient
to provide the normal heat conductivity. However, the latter study confirms the convergence for
the models which, similarly to the LJ, have bounded attractive branch of the nearest-neighbor
interaction potential.
In other terms, this family of models allows "dissociation"\  of the
neighboring particles and formation of effective "gaps"\  in the chain. It was conjectured that
these gaps are responsible for convergence of the heat conduction.
This conjecture is a main subject of current paper.

In generic nonlinear potential it seems very difficult to distinguish between various factors
which effect the heat conductivity, and to filter out the effects related to a topology of the potential function.
To overcome this difficulty, we consider a chain of elastic disks. In the ground states for
high density this system is in fact a linear chain. So, the gap formation is the only possible
mechanism of the phonon scattering. Thus, this model allows one to study the role
of the dissociation directly.

The model chain comprises $N$ one-dimensional ``disks'' with
elastic compressive nearest-neighbor interaction. In other terms,
the system Hamiltonian is expressed as
%--------------------------------- 1-----------------------------------
\begin{equation}
\mathrm{\mathit{\mathcal{H}}}=\sum_{n=1}^{N}p_{n}^{2}/2M+\sum_{n=1}^{N-1}U(x_{n+1}-x_{n}),
\label{f1}
\end{equation}
%--------------------------------- 1-----------------------------------
where M is the mass of each particle, $x_{n}$ is a coordinate of the
center of $n$-th disk, $p_{n}$ -- its momentum. The potential of interaction
in this model is defined as:
%---------------------------- 2--------------------------------
\begin{equation}
 U(r)= \left\{
 \begin{array}{ll}
 K(r-D)^{2}/2, & r<D \\
 0, & r\geq D
 \end{array}
 \right.,
\label{f2}
%---------------------------- 2 ------------------------------
\end{equation}
where $K$ -- stiffness of the disks and $D$ -- their diameter.
The neighboring disks repulse each other linearly, when their centers
are at a distance less than $D$. In order to keep a "numeric"\ density
(a number of the disks at unit length) constant,
we adopt that terminal disks are fixed: $x_{1}(t)\equiv 0$, $x_{N}(t)\equiv (N-1)a$.
Here $a$ is average length of the link between the neighboring particles.

Introducing dimensionless time $\tau=t\sqrt{K/M}$, displacements
$u_{n}=x_{n}/a$ and energy $H=\mathcal{H}/Ka^{2}$, we arrive to
the following non-dimensional Hamiltonian:
%--------------------------- 3 --------------------------------
\begin{equation}
H=\sum_{n=1}^{N}{u_{n}'}^2/2+\sum_{n=1}^{N-1}V(u_{n+1}-u_{n}),
\label{f3}
\end{equation}
%--------------------------- 3 -------------------------------
where apostrophe denotes differentiation with respect to $\tau$.
Non-dimensional potential of interaction is described as
%--------------------------- 4 -------------------------------
\begin{equation}
V(\rho)= \left\{
\begin{array}{ll}
(\rho-d)^{2}/2, & \rho<d \\
             0, & \rho\geq d
\end{array}
\right. .
\label{f4}
\end{equation}
%--------------------------- 4 -------------------------------
Parameter $d=D/a$ is related to a "packing density" of the disks
(numeric density is always unit, as it was explained above).

\begin{figure}[t,b]
\includegraphics[angle=0, width=1\linewidth]{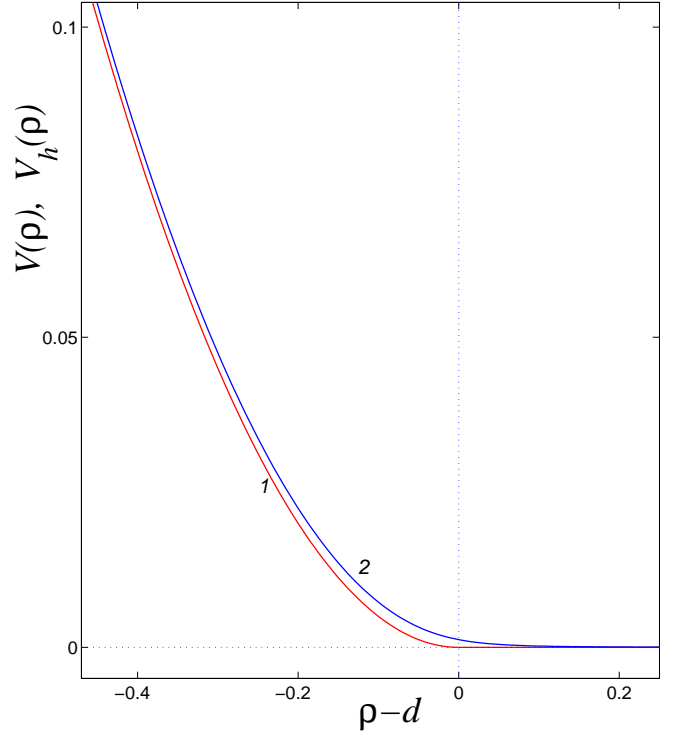}
\caption{(Color online)
Comparison between the potential $V(\rho)$ (curve 1)
and smoothened potential $V_h(\rho)$ with $h=0.01$ (curve 2).
}
\label{fig1}
\end{figure}

Potential (\ref{f4}) has a discontinuity
of the second derivative. To avoid numeric complications, we approximate
it by smoothened potential in a form
%-------------------------- 5 ----------------------------
\begin{equation}
V_h(\rho)=\left[\sqrt{(\rho-d)^{2}+h}-(\rho-d)\right]^{2}/8.
\label{f5}
%-------------------------- 5 -----------------------------
\end{equation}
Value of parameter $h$ determines the accuracy of the smoothening. A limit
$h\rightarrow 0$ corresponds to non-smooth potential (\ref{f4}).
For simulations at relatively high temperatures $T\ge 0.09$ we use $h=0.01$
and for lower temperatures -- $h=0.0005$. The quality of approximation (\ref{f5})
for higher value of $h$ is illustrated in Fig.~\ref{fig1}.
%---------------------------- figure 1 ----------------------------
%---------------------------- figure 1 ----------------------------

In order to simulate the heat transfer in this model, we used Langevin
stochastic thermostat. To verify the results, we used both commonly accepted
methods for computation of the heat conduction coefficient --- based
on stationary heat transfer and on linear response theory.
To simulate the stationary heat transfer, we considered the chain with Hamiltonian
(\ref{f3}) and potential (\ref{f5}), comprising in general $N_{+}+N+N_{-}$ disks.
Terminal fragments of the chain with $N_{+}$ and $N_{-}$ particles are permanently attached
to "hot"\  and "cold"\  thermostats respectively.
Temperature gradient and heat flux are measured at free middle part of the chain;
this part comprises $N$ particles.
To be even more specific, we simulated numerically the following system of dynamic equations:
%------------------------------ 6 -----------------------------------
\begin{eqnarray}
u_{n}^{\prime\prime} & = & -\partial H/\partial u_{n}-\gamma u_{n}^{\prime}+\xi_{n}^{+},~
1<n\leq N_{_{+}},\nonumber \\
u_{n}^{\prime\prime} & = & -\partial H/\partial u_{n},~
N_{+}<n\leq N_{_{+}}+N, \label{f6}\\
u_{n}^{\prime\prime} & = & -\partial H/\partial u_{n}-\gamma u_{n}^{\prime}+\xi_{n}^{-},
\nonumber\\
&& N_{+}+N<n<N_++N+N_-.\nonumber
\end{eqnarray}
%-------------------------------- 6 -----------------------------------
Here $\gamma$ is a relaxation coefficient, $\xi_{n}^{\pm}$ is a white
Gaussian noise, normalized by the following conditions:
\begin{eqnarray}
\left\langle \xi_{n}^{\pm}(\tau)\right\rangle =0,~~
\left\langle \xi_{n}^{+}(\tau_{1})\xi_{k}^{-}(\tau_{2})\right\rangle =0, \nonumber \\
\left\langle \xi_{n}^{\pm}(\tau_{1})\xi_{k}^{\pm}(\tau_{2})\right\rangle
=2\gamma T_{\pm}\delta_{nk}\delta(\tau_{1}-\tau_{2}). \nonumber
\end{eqnarray}
%------------------------------ Fig. 2 --------------------------------------
\begin{figure}[tb]
\includegraphics[angle=0, width=1\linewidth]{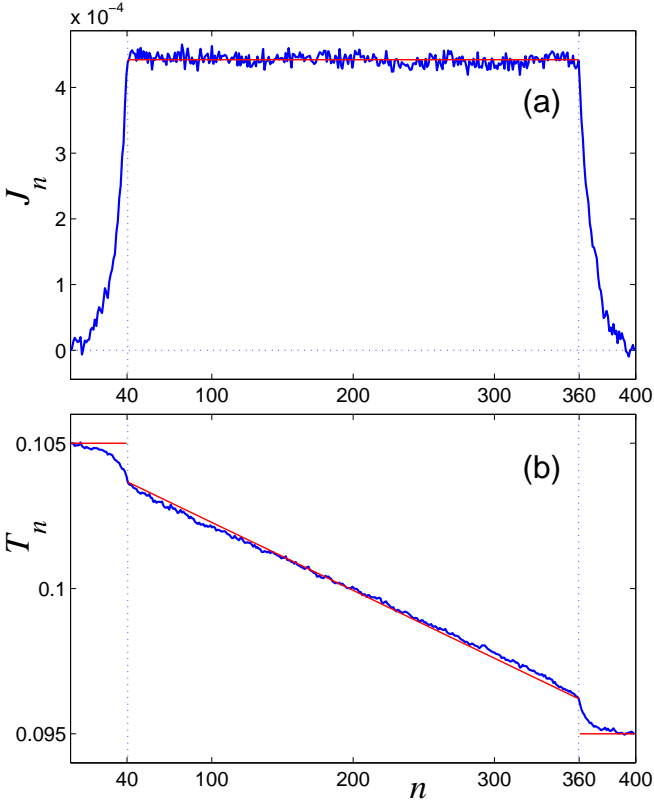}
\caption{(Color online)
Distribution of (a) average heat flux $J_n$ and (b) local temperature $T_n$ in the chain
of elastic disks with $d=1$ and $N_\pm=40$, $N=320$, $T_+=0.105$, $T_-=0.095$.
Straight red line approximates the linear temperature gradient which was used
for evaluation of the heat conduction coefficient $\kappa(N)$.
}
\label{fig2}
\end{figure}
%------------------------------ Fig. 2 --------------------------------------

System of equations (\ref{f6}) was integrated numerically by Velocity Verlet method.
After initial transient, stationary heat flux at free part of the chain has been observed.
Then, we computed the values of average local temperature defined as
$T_{n}=\left\langle u_{n}^{\prime2}\right\rangle _{\tau}$
and of average local heat flux $J_{n}=\left\langle j_{n}\right\rangle_{\tau}$,
where $j_{n}$ denotes an instantaneous local heat flux:
$$
j_{n}=(u_{n+1}-u_{n})(u_{n}^{\prime}+u_{n+1}^{\prime})F(u_{n+1}-u_{n})/2+u_{n}^{\prime}h_{n},
$$
$F(\rho)=-\partial V_h/\partial\rho$ and $h_{n}$ is a local energy density:
$$
h_{n}=[u_{n}^{\prime2}+V_h(u_{n+1}-u_{n})+V_h(u_{n}-u_{n-1})]/2
$$
(see \cite{LLP03}). All time averages were defined as
$$
\left\langle f\right\rangle _{\tau}=\lim_{\tau\rightarrow\infty}\tau^{-1}\int_{0}^{\tau}f(\tau)d\tau.
$$
In all simulations, we will use the following parameter values:
$T_{\pm}=(1\pm0.05)T$, $\gamma=0.05$, $N_{\pm}=40$,
$N=20$, 40, ...,  2560.
%---------------------- Fig. 3 ------------------------------
\begin{figure}[tb]
\includegraphics[angle=0, width=1\linewidth]{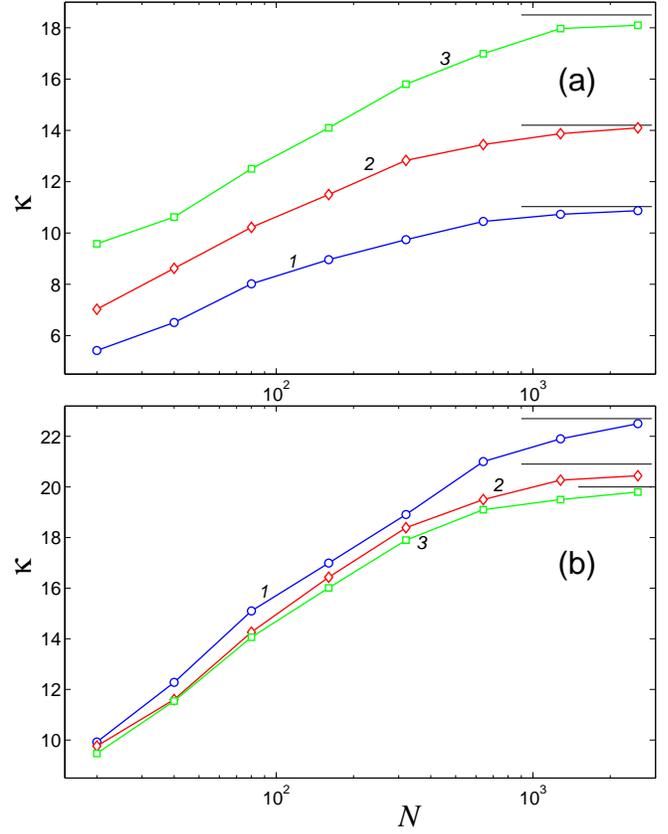}
\caption{(Color online)
Dependence of heat conduction coefficient $\kappa$ on the length of free internal chain fragment
$N$ for the model of elastic disks with (a) $d=0.1$ and (b) $d=1$ for
temperatures $T=0.1$, 1, 10 (curves 1, 2, 3 respectively).
Horizontal lines correspond to the values of the heat conduction coefficient obtained
with help of Green-Kubo formula (\ref{f8}).
}
\label{fig3}
\end{figure}
%---------------------- Fig. 3 ------------------------------

As it is illustrated in Fig.~\ref{fig2}, the heat flux $J=J_{n}$  through the central
(free) segment of the chain was stationary (cite-independent).
The heat conduction coefficient has been then evaluated as
%--------------------------------------- 7 ----------------------------------
\begin{equation}
\kappa(N)=J(N-1)/(T_{N_{+}+1}-T_{N_{+}+N}).
\label{f7}
\end{equation}
%--------------------------------------- 7 -----------------------------------

Well-known alternative way to evaluate the heat conduction coefficient is based
on well-known Green-Kubo formula
%-------------------------------- 8------------------------
\begin{equation}
\kappa_{c}=\lim_{\tau\rightarrow\infty}\lim_{N\rightarrow\infty}\frac{1}{NT^{2}}\int_{0}^{\tau}c(s)ds,
\label{f8}
\end{equation}
%-------------------------------- 8 ------------------------
where $c(s)=\left\langle J_{tot}(\tau)J_{tot}(\tau-s)\right\rangle _{\tau}$
is autocorrelation function of the total heat flux in the
system $J_{tot}(\tau)=\sum_{n=1}^{N}j_{n}(\tau)$.

In order to compute the autocorrelation function $c(\tau)$ we consider
a cyclic chain consisting of $N=10^{4}$ particles. Initially all particles in this chain are
coupled to the Langevin thermostat with temperature $T$.
After achieving the thermal equilibrium, the system is detached from
the thermostat and Hamiltonian dynamics is simulated. To improve the
accuracy, the results were averaged over $10^4$ realizations of
the initial thermal distribution.

Both tests indicate that the heat conduction coefficient of the system
converges in the thermodynamical limit for all studied values of the
parameters. The results are summarized in Fig.~\ref{fig3}.

\begin{figure}[tb]
\includegraphics[angle=0, width=1\linewidth]{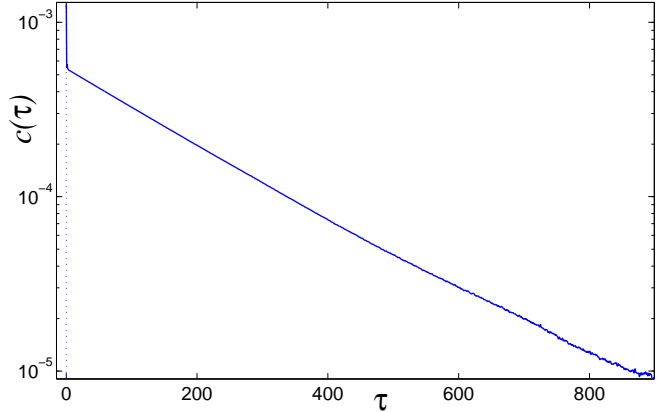}
\caption{
Exponential decay of the autocorrelation function $c(\tau)$ for the chain
of elastic disks with $d=0.1$ and temperature $T=0.1$.
}
\label{fig4}
\end{figure}

Horizontal lines at these figures correspond to the values computed from Green-Kubo
formula (\ref{f8}). The latter simulation also reveals exponential decay
of the autocorrelation function, sufficient for convergence of the
integral (\ref{f8}). This result is illustrated in Fig.~\ref{fig4}.
Both tests yield very close results for large values of $N$.
%(note logarithmic scale of $N$ axis in Fig.~\ref{fig3}).
%------------------- Fig. 4 -------------------------------------
%-------------------- Fig. 4 ------------------------------------

Our next goal is to explain the above findings. First, let us
consider the regime of dense packing, in which $d>1$.
Since the numeric density of the considered chain
is always unit, at zero temperature all disks are compressed and the
system corresponds to common linear homogeneous chain. However, as
the temperature increases, there exists certain probability to form a
gap between the neighboring particles due to thermal fluctuations.
Scattering at these gaps is the only possible mechanism to provide
the phonon scattering, since in the fragments without the gaps
the oscillatory waves do not interact at all. Thus, it is reasonable
to conjecture, that a mean free path is governed by simple Arrhenius
formula. The velocity of the oscillatory waves can be estimated as
a sound velocity and is equal to unity in our non-dimensional model. Then,
the heat conduction coefficient may be crudely estimated as:
%--------------------------- 9 ---------------------------
\begin{equation}
\kappa\sim\exp[\alpha(d-1)^{2}/T].
\label{f9}
\end{equation}
%--------------------------- 9 ---------------------------

\begin{figure}[t]
\includegraphics[angle=0, width=1\linewidth]{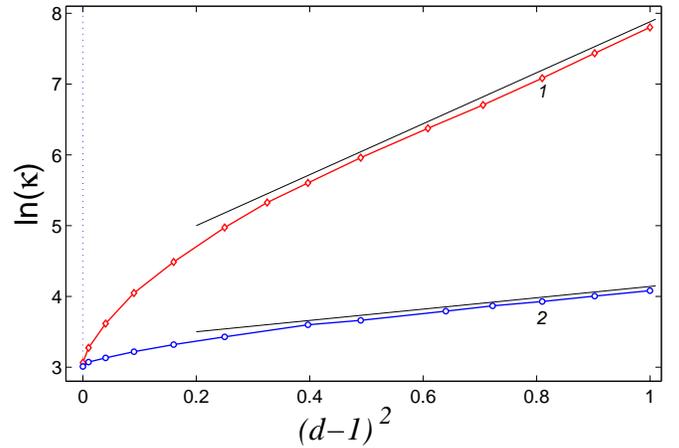}
\caption{(Color online)
Dependence of $\ln(\kappa)$ on $(d-1)^2$ for average temperatures $T=0.1$, 1 (curves 1, 2 respectively).
Only values $d>1$ are presented.
Straight lines give linear dependence for relatively large densities
}
\label{fig5}
\end{figure}

\begin{figure}[t]
\includegraphics[angle=0, width=1\linewidth]{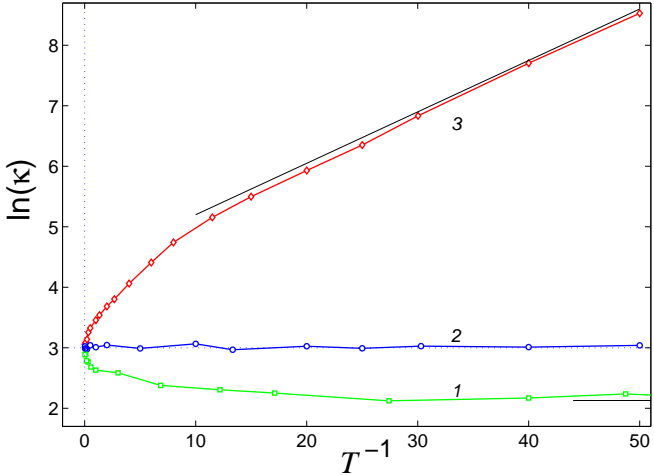}
\caption{(Color online)
Dependence of $\ln(\kappa)$ on inverse temperature $T^{-1}$ for the disk diameters $d=0.1$, $d=1$
and $d=1.5$ (curves 1, 2, 3 respectively).
Upper straight line gives Arrhenius scaling for case $d=1.5$.
Lower straight line gives value of heat conductivity in limit
$T\rightarrow 0$ for case $d=0.1$.
}
\label{fig6}
\end{figure}

Scaling of $\ln\kappa$ with $(d-1)^2$ and $T^{-1}$ is presented in Figs.~\ref{fig5} and \ref{fig6}.
One can see that the results nicely correspond to estimation
(\ref{f9}) for $d>1$ and relatively large temperatures. This
result confirms our understanding of the reasons for normal heat conductivity in this system.
%--------------------- Fig. 5 ----------------------------
%--------------------- Fig. 5 -------------------------------
%--------------------- Fig. 6 -------------------------------
%--------------------- Fig. 6 -------------------------------

The case of loose packing of the disks, corresponding to $d< 1$, also deserves special attention.
As the temperature
is relatively small, the disks will move free for most of the time.
Due to linear elastic interaction the model, the time of each binary
collision $\tau_{c}$ between the disks will not depend on the temperature
and the density: $\tau_{c}=\pi$. The collisions are elastic.
If the collisions would be only binary, the momentum of colliding particles
would be simply exchanged in each collision.
Therefore, the system would be equivalent
to a billiard of particles with equal masses on a line, and therefore would be
completely integrable. Thus, finite heat conduction coefficient is intimately related
to triple (and higher-order) collisions.
One can adopt (in a limit of very loose packing, $d << 1$ ) that a mean free path $\lambda_{imp}$
of an impulsive excitation transferred through the binary collisions
is inversely proportional to a probability that the disk will hit
a system of two already interacting disks. This probability can be estimated
as the ratio of the times of the interaction and the free flight.
The latter, in turn, can be estimated as $\tau_{0}=(1-d)/\sqrt{T}$
(average velocity of the disk $v\sim\sqrt{T}$). Then, the heat
conduction coefficient for small temperatures and small $d$ may be
crudely estimated as:
%-------------------------------- 10 ---------------------------------
\begin{equation}
\kappa\sim v\lambda_{imp}\sim v\tau_{0}/\tau_{c}\sim(1-d)/\pi.
\label{f10}
\end{equation}
%-------------------------------- 10 ---------------------------------

Equation (\ref{f10}) predicts that the heat conduction coefficient
will not depend on the temperature in the limit of loose packing. Numeric
simulation completely confirms this prediction -- see Fig. \ref{fig6}.

The chain with $d=1$ is an intermediate case between the two limits
described above. Formation of the oscillatory waves is hardly possible,
since the energy required for the gap formation is zero. From the
other side, the free motion of the disk is also largely suppressed.
In this case, as it seems, the heat conduction coefficient will weakly
depend on the temperature, as it is illustrated in Fig. \ref{fig6}. The value
of the heat conduction coefficient in this case is close to $\kappa=20$
in all temperature diapason. The heat conductivity in similar model
for this degenerate case has been presented in paper \cite{1dElastic}. There
they reported the value $\kappa=2$. It seems that this discrepancy
is related to Nose-Hoover thermostats used in \cite{1dElastic} for the noon-equilibrium
simulations. As it is well-known, this choice of the thermostats can
easily lead to artifacts.

The chain of elastic disks is in a sense unique model -- the dissociation is the only possible
mechanism of the phonon scattering. However, in more complicated nonlinear chains the role of the
dissociation might be less profound, since there are other mechanisms of the phonon scattering
-- just due to the nonlinearity of the interaction potential. To shed a light on a role
of the possibility of dissociation, we simulate the heat
conduction in the Lennard-Jones chain with the nearest-neighbor potential
%-------------------------- 11 -------------------------------
\begin{equation}
U(x_{n+1}-x_{n})=4\varepsilon[(\sigma/(x_{n+1}-x_{n})^6-1/2]^2
\label{f11}
\end{equation}
%-------------------------- 11 -------------------------------
with $\sigma=2^{-1/6}$ and energy of coupling $\varepsilon=1/72$.

\begin{figure}[t]
\includegraphics[angle=0, width=1\linewidth]{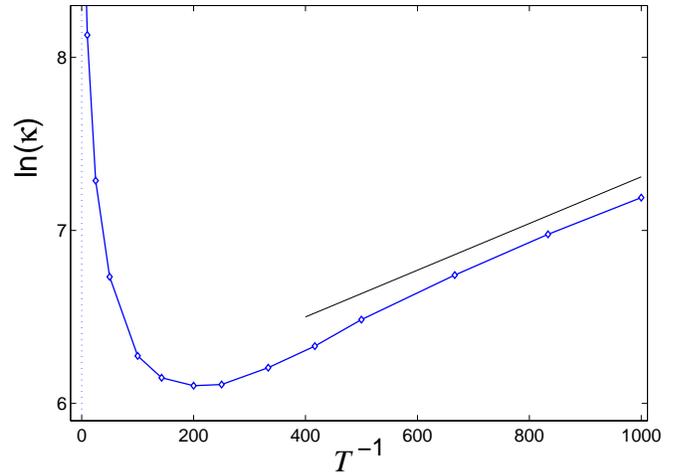}
\caption{
Dependence of $\ln\kappa$ on inverse temperature $1/T$ for the LJ chain.
Straight line corresponds to Arrhenius-like behavior for low temperatures.
}
\label{fig7}
\end{figure}

The stationary heat transfer has been simulated according to the protocol described above,
with $N_{\pm}=40$, $N=2560$, $T_{\pm}=T(1\pm 0.1)$.
Results of this simulation are presented in Fig. \ref{fig7}.
%------------------------- Fig. 7 -------------------------------
%------------------------- Fig. 7 ------------------------------

These results demonstrate Arrhenius-type behavior of the heat conduction coefficient
in the limit of low temperatures also for the LJ chain. In the chain of elastic disks such behavior
may be attributed to the phonon scattering related to the dissociation.
Thus, one can conjecture that also in the LJ chain the convergence of the heat conduction
coefficient is related to the possibility of dissociation. In the limit of high temperatures
the heat conduction coefficient sharply increases. This observation can be related to strong
repulsive core of the LJ potential, which plays main role in the case of relatively high temperatures.

To conclude, we demonstrate that finite coupling energy of the neighboring particles in
the model chain can play decisive role in providing the convergent heat conduction. This statement
is completely clarified for simple model of linearly elastic disks, and obtain certain support
for the model of LJ particles, at least for the case of low temperatures.
Notably, very simple considerations related to the mean free path and average velocity
of elementary excitations, are sufficient to describe the asymptotic
behavior of the heat conduction coefficient. It seems the nonlinearity of the potential,
related to the possibility of dissociation,
is "strong"\  enough to make the heat conduction to conform to these well-known models.

\acknowledgments
The authors are grateful to Israel Science Foundation (grant 838/13)
for financial support. A. V. S. is grateful to the Joint Supercomputer Center
of the Russian Academy of Sciences for the use of computer facilities.

%-----------------------------------------------------------------------------------------------------

%end{references}
\end{document}